\documentclass[twocolumn,english,aps,superscriptaddress,pre,longbibliography]{revtex4-1}
\usepackage[T1]{fontenc}
\usepackage[latin9]{inputenc}
\usepackage{color}
\usepackage{babel}
\usepackage{amsmath}
\usepackage{amssymb}
\usepackage{graphicx}
\usepackage{esint}
\usepackage[unicode=true,pdfusetitle,
 bookmarks=true,bookmarksnumbered=false,bookmarksopen=false,
 breaklinks=true,pdfborder={0 0 0},backref=false,colorlinks=true]
 {hyperref}
\hypersetup{
 linkcolor=blue,citecolor=blue,urlcolor=blue}

\makeatletter

\providecommand{\tabularnewline}{\\}

\usepackage{pslatex}

\usepackage{color}
\usepackage{babel}

\usepackage{subfigure}
\makeatother

\begin{document}

\title{Percolation on Isotropically Directed Lattice}

\author{Aurelio W. T. de Noronha}

\affiliation{Departamento de Física, Universidade Federal do Ceará, 60451-970 Fortaleza, Ceará, Brazil }

\author{André A. Moreira}

\affiliation{Departamento de Física, Universidade Federal do Ceará, 60451-970 Fortaleza, Ceará, Brazil }

\author{André P. Vieira}

\affiliation{Instituto de Física, Universidade de Sao Paulo, Caixa Postal 66318, 05314-970, São Paulo, Brazil}
\affiliation{Computational Physics for Engineering Materials, IfB, ETH Zurich, Schafmattstr. 6, CH-8093 Z Zürich, Switzerland }

\author{Hans J. Herrmann}

\affiliation{Departamento de Física, Universidade Federal do Ceará, 60451-970 Fortaleza, Ceará, Brazil }
\affiliation{Computational Physics for Engineering Materials, IfB, ETH Zurich, Schafmattstr. 6, CH-8093 Z Zürich, Switzerland }

\author{José S. Andrade Jr}

\affiliation{Departamento de Física, Universidade Federal do Ceará, 60451-970 Fortaleza, Ceará, Brazil }

\author{Humberto A. Carmona}

\affiliation{Departamento de Física, Universidade Federal do Ceará, 60451-970 Fortaleza, Ceará, Brazil }

\date{\today}
\begin{abstract}
We investigate percolation on a randomly directed lattice, an
intermediate between standard percolation and directed percolation,
focusing on the isotropic case in which bonds on opposite directions
occur with the same probability. We derive exact results for the
percolation threshold on planar lattices, and present a conjecture for
the value the percolation-threshold for in any lattice. We also
identify presumably universal critical exponents, including a fractal
dimension, associated with the strongly-connected components both for planar
and cubic lattices.  These critical exponents are different from those
associated either with standard percolation or with directed
percolation.
\end{abstract}
\maketitle

\section{Introduction}

In a seminal paper published some 60 years ago, Broadbent and
Hammersley \citep{Broadbent1957} introduced the percolation model, in
a very general fashion, as consisting of a number of sites
interconnected by one or two directed bonds which could transmit
information in opposite directions. However, over the years, most of
the attention has been focused on the limiting cases of standard
percolation, in which bonds in both directions are either present or
absent simultaneously, and of directed percolation, in which only
bonds in a preferred direction are allowed. 
While standard percolation represents one of the simplest models for
investigating critical phenomena in equilibrium statistical physics~\citep{Stauffer1994}, directed percolation has become a paradigmatic
model for investigating non equilibrium phase
transitions~\citep{Marro2005}. Moreover, it has been shown that the
isotropic case, where bonds in both opposite directions are present
with the same probability is a very particular case, with any amount
of anisotropy driving the system into the same universality class as
that of directed percolation~\citep{Janssen2000,Hu2014,Redner1982}.

The case of percolation on isotropically directed lattices has
received much less attention. This modified percolation model should
be particularly relevant to the understanding of a large number of
physical systems. For instance, in the same way that standard
percolation was shown to be related to other models in statistical
mechanics~\citep{Fortuin1972} one could expect percolation on
isotropically directed lattices to be related to statistical systems
with non-symmetric interactions~\citep{Lima2010}. It has been shown
that identifying the connected components systems with non-symmetric
interactions can elucidate questions regarding the
controllability~\citep{Liu2016} and observability~\citep{Santolini2018} 
of these systems. Percolation with directed bonds have also been 
investigated in the field of traffic dynamics~\citep{Li2015}.Redner~\citep{Redner1981,Redner1982,Redner1982b} formulated the problem of percolation on isotropically directed
lattices as a random insulator-resistor-diode circuit model, in which
single directed bonds represent diodes, allowing current to flow in
only one direction, while double bonds in opposite directions
represent resistors and absent bonds represent insulators.

Focusing on hypercubic lattices, he employed an approximate real-space
renormalization-group treatment which produces fixed points associated
with both standard percolation (in which only resistors and insulators
are allowed) and directed percolation (in which only insulators and
diodes conducting in a single allowed direction are present), as well
as other ``mixed'' fixed points controlling lines of critical points,
for cases in which the all three types of circuit elements are
present. The crossover from isotropic to directed percolation when
there is a slight preference for one direction was studied via
computer simulations \citep{Inui1999} and renormalized field theory
\citep{Janssen2000,Stenull2001}. More recently, the same crossover
problem was independently investigated on the square and on the
simple-cubic lattices by Zhou et al. \citep{Zhou2012}, who dubbed
their model ``biased directed percolation''.

\begin{figure}
\includegraphics[width=0.40\textwidth]{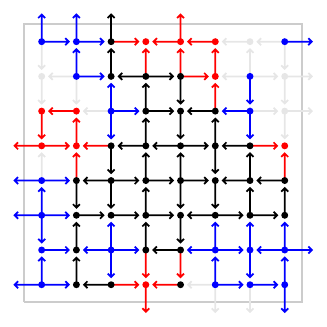}
\caption{A square lattice where each pair of nearest neighbors
  is connected by a directed bond at the critical point ~\cite{Zhou2012}. The color code is as follows:
  Black indicates nodes and bonds comprising the largest strongly connected component (GSCC), that is,
  the largest set of nodes that can be mutally reached from each other.
  Red indicates nodes and bonds outside the GSCC that can be reached from nodes in the GSCC.
  Blue indicates nodes and bonds outside the GSCC from where nodes in the GSCC can be reached. The largest outgoing/incoming  component (GOUT/GIN) includes all nodes of the GSCC augmented by the
  red/blue nodes, respectively. Grey indicates nodes and bonds outside both GIN and GOUT.
  Bonds exiting the enclosing box represent connections through the periodic boundary condition.
}
\label{lattice}
\end{figure}

We are interested here in percolation of isotropically  directed
bonds in which bonds in opposite
directions are present with the same probability, possibly along with
vacancies and undirected bonds. It has been conjectured that this
model is in the same universality class as standard
percolation~\cite{Janssen2000,Zhou2012}, however these works have
focused on the sets of nodes that can be reached from a given point.
In fact, when considering directed bonds, it is possible that site A
can be reached from site B, while site B cannot be reached from site
A, what therefore calls for a redefinition of a cluster. Percolation
of directed bonds was investigated within the context of complex
networks~\citep{Dorogovtsev2001,Schwartz2002,Boguna2005,Kenah2007,Serrano2007,Frasceschet2012,Zhu2014,Li2015},
where the concept of strongly-connected components (SCCs) has been
adopted~\citep{Tarjan1972}, defined as those sets of points which can
be mutually reached following strictly the bond directions. A critical
state of the model can be characterized as the point where a giant
strongly connected component (GSCC) is
formed~\citep{Dorogovtsev2001}. Alternatively, one can define a giant
cluster formed by all the sites that can be reached from a given site
following bond directions (GOUT)~\citep{Dorogovtsev2001}, and
determine the critical point where such cluster is formed.
Alternatively, one can define a giant
cluster formed by all the sites that can be reached from a given site
following bond directions~\citep{Dorogovtsev2001}, and determine the
critical point where such cluster is formed. There is no
logical need for these two points to be the same, leaving the
possibility of two distinct phase transitions existing in this model
\citep{Schwartz2002}. However, both for regular lattices, as we will show
here, and for some complex networks~\citep{Dorogovtsev2001},
these two objects form at the same critical point. In
Fig.~\ref{lattice} we show an example of a square lattice at the
critical point.

This paper is organized as follows. In Section II we define the model and
present calculations of percolation thresholds. In Section III we discuss
some exact results on hierarchical lattice that shed light
on the critical state of this model.  Our computer simulation results
are presented in Section IV, while Section V is dedicated to a concluding
discussion.

\section{Definition of the model and calculation of percolation thresholds}

We work on $d$-dimensional regular lattices. All sites are assumed to
be present, but there are a few possibilities for the connectivity
between nearest-neighbors. With probability $p_{0}$ they may not be
connected (indicating a vacancy). With probability $p_{1}$ they may be connected by a directed bond (with equal probabilities for either
direction). Finally, with probability $p_{2}$ neighbors may be
connected by an undirected bond (or equivalently by two having opposite directions). Of course we must fulfill $p_{0}+p_{1}+p_{2}=1$.

A simple heuristic argument yields an expression for the critical
threshold for percolation of isotropically directed bonds. Starting from a
given site $i$ on a very large lattice, the probability
$p_{nn}$ that a particular nearest-neighbor site can be reached
from $i$ is given by the probability that both directed bonds are
present between these neighbors ($p_{2}$) plus the probability that
there is only one directed bond and that it is oriented in the
appropriate direction ($\frac{1}{2}p_{1}$). As the distribution of orientations is on
average isotropic, the critical threshold must depend only on
$p_{nn}$.  In fact, using the Leath-Alexandrowicz method
~\citep{Leath1976,Alexandrowicz1980}, it can be
shown~\citep{Zhou2012} that the clusters of sites reached from a seed
site in percolation of directed bonds with a given $p_{nn}$ are
identically distributed to the clusters of standard percolation with an
occupation $p_{sp}$, as long as $p_{sp}=p_{nn}$. Therefore,
we conclude that the critical percolation probabilities of our model should fulfill
\begin{equation}
p_{2}+\tfrac{1}{2}p_{1}=p_{c},\label{eq:pc}
\end{equation}
in which $p_{c}$ is the bond-percolation threshold for standard percolation
in the lattice.

\begin{figure}
\subfigure[]{\includegraphics[width=1\columnwidth]{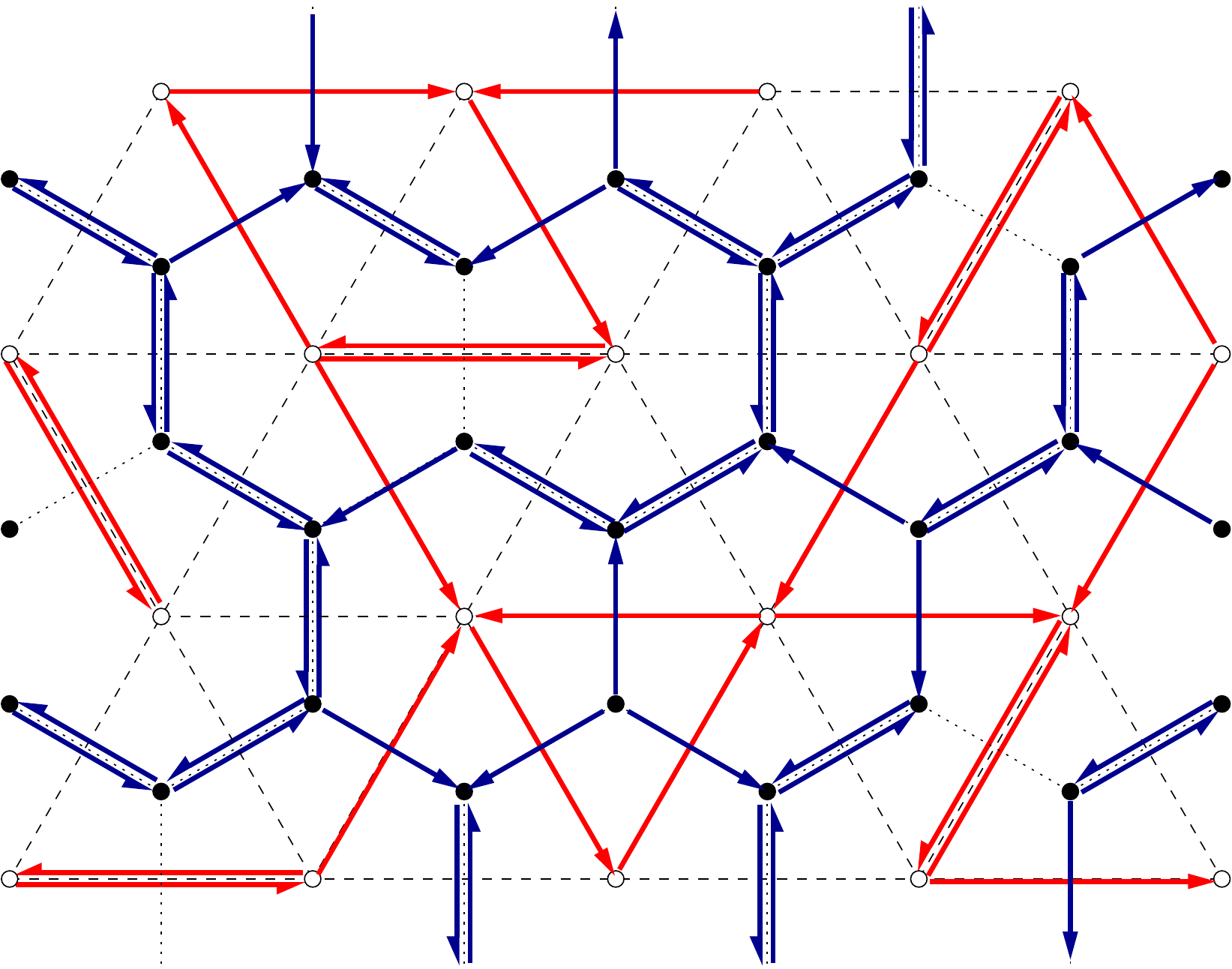}}\linebreak{}
\subfigure[]{\includegraphics[width=0.5\columnwidth]{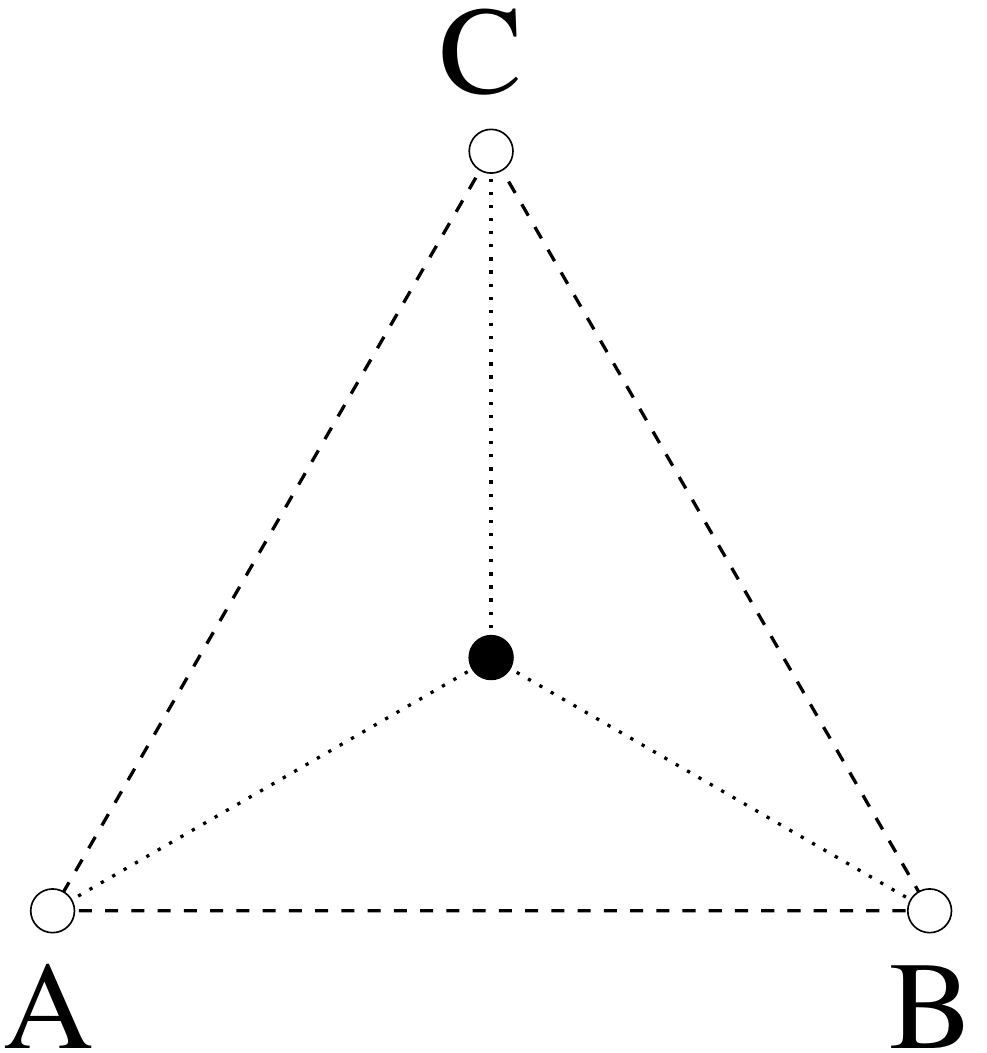}
}

\caption{\label{fig:startriangle}(a) Illustration of a configuration
  of percolation of directed bonds on the triangular (white sites, red
  arrows) and honeycomb (black sites, blue arrows) lattices, related
  by the dual transformation defined in the text.
  (b) The sites involved in the star-triangle transformation discussed
  in the text. }
\end{figure}

We can use duality arguments to show that Eq. (\ref{eq:pc}) is indeed
exact for the square, triangular and honeycomb lattices. A duality
transformation for percolation of directed bonds on planar lattices
was previously introduced~\citep{Redner1982} to derive the percolation
threshold on the square lattice. The transformation states that every
time a directed bond is present in the original lattice, the directed
bond in the dual lattice that crosses the original bond forming an
angle of $\frac{\pi}{2}$ clockwise will be absent. With the opposite
also holding, namely every time a directed bond is absent in the original
lattice, in the dual lattice the bond forming a angle $\frac{\pi}{2}$
clockwise will be present.  Of course, an undirected bond (or
alternatively two bonds in opposite directions) in the original
lattice corresponds to a vacancy in the dual lattice, and
vice-versa. Figure~\ref{fig:startriangle}(a) shows a configuration of
percolation of directed bonds on the triangular lattice and the
corresponding dual honeycomb lattice.

Denoting by $q_{0}$, $q_{1}$ and
$q_{2}$ the respective probabilities that there is a vacancy, a single directed
bond, or an undirected bond between nearest neighbors on the dual lattice, the transformation
allows us to write,
\begin{equation}
q_{0}=p_{2},\quad q_{1}=p_{1},\quad q_{2}=p_{0}.\label{eq:dualtransf}
\end{equation}
From these results and the normalization conditions 
\begin{equation}
p_{0}+p_{1}+p_{2}=q_{0}+q_{1}+q_{2}=1\label{eq:normal}
\end{equation}
 we immediately obtain
\[
\tfrac{1}{2}\left(p_{2}+q_{2}\right)+\tfrac{1}{2}p_{1}=\tfrac{1}{2},
\]
which is valid for any choice of the probabilities. As already noticed
by Redner \citep{Redner1982}, for the square lattice, which is its
own dual, we must have $p_{2}=q_{2}$ at the percolation threshold,
yielding
\begin{equation}
p_{2}+\tfrac{1}{2}p_{1}=\tfrac{1}{2},\label{eq:pcsquare}
\end{equation}
in agreement with Eq. (\ref{eq:pc}). Here, of course, we assume that there
is only one critical point.

\begin{figure}
	\includegraphics[width=0.40\textwidth]{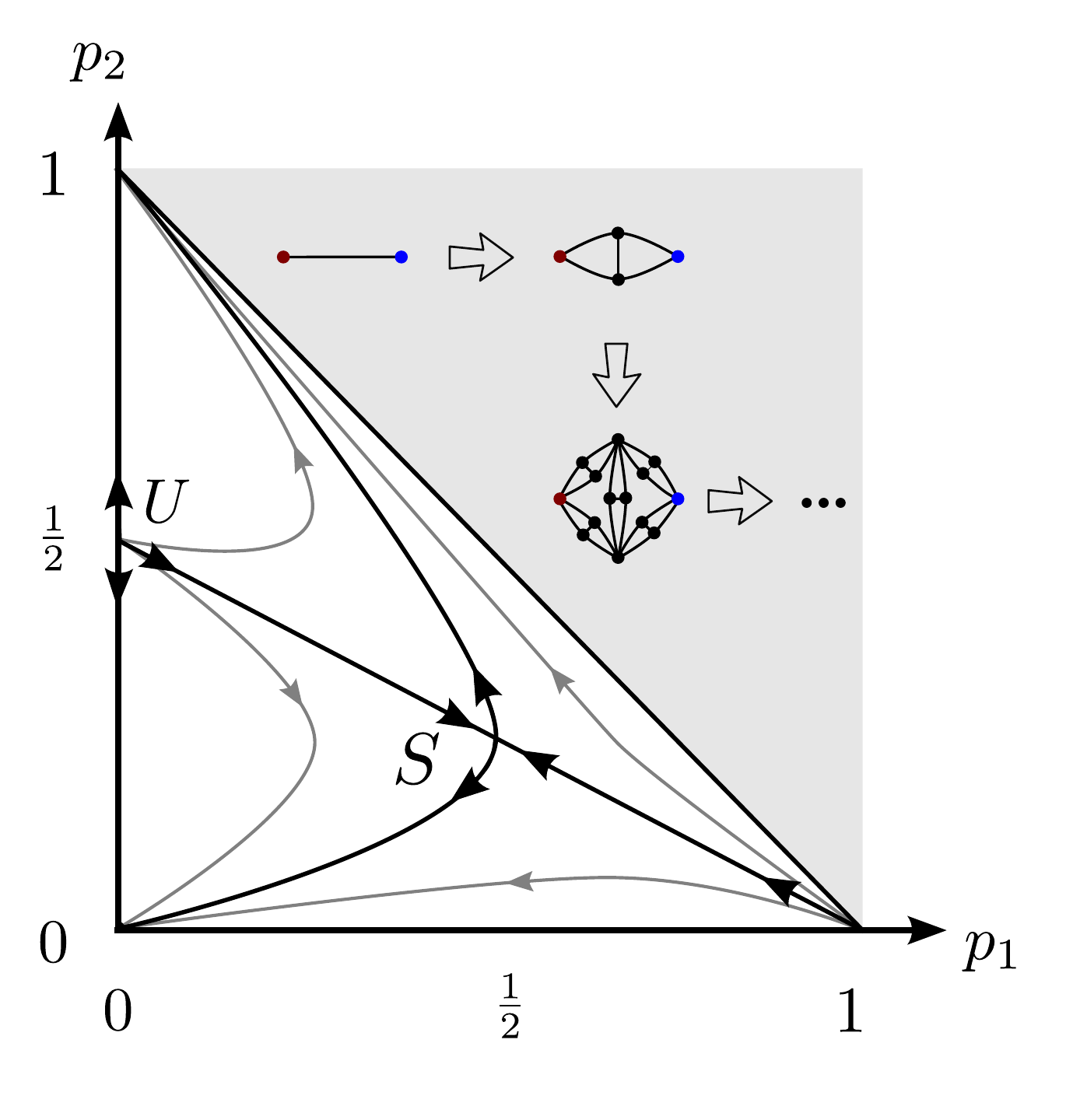}
	\caption{Phase diagram for percolation of isotropically directed
          bonds on the hierarchical lattice obtained as the limit of
          the process displayed. Redner~\citep{Redner1981,Redner1982}
          and Dorogovtsev~\citep{Dorogovtsev1982} used the renormalization
          group to solve exactly this model on the shown hierarchical
          lattice. The directions of the lines indicate the
          renormalization group flux.  The critical line
          $p_2+p_1/2=1/2$ coincides with the critical line for percolation of isotropically directed bonds on the square
          lattice. The renormalization group shows that any point on the critical line, besides $U=(0,1/2)$ that corresponds to
          the critical point of standard percolation, will display the
          same scale invariant behavior as a point $S$ located along
          the critical line, suggesting the possibility that percolation of isotropically directed bonds may be in a different
          universality class from standard percolation.}
  \label{diagrama}
\end{figure}

The triangular and the honeycomb lattices are related by the duality
transformation, as illustrated in Fig. \ref{fig:startriangle}(a),
and we now use a star-triangle transformation \citep{Sykes1964} to
calculate their bond-percolation thresholds. Based on enumerating
the configurations of bonds connecting the sites identified in Fig.
\ref{fig:startriangle}(b), we can calculate the probabilities $P$
and $Q$ of connections between the sites on the star and on the triangle, respectively. For the probability that site A is connected only to site
B or only to site C, we obtain
\[
P_{\text{AB}}=\tfrac{1}{2}p_{1}p_{2}^{2}+\tfrac{1}{2}p_{1}^{2}p_{2}+\tfrac{1}{8}p_{1}^{3}+p_{0}p_{2}^{2}+p_{0}p_{1}p_{2}+\tfrac{1}{4}p_{0}p_{1}^{2},
\]
\[
Q_{\text{AB}}=\tfrac{1}{4}q_{1}^{2}q_{2}+\tfrac{1}{8}q_{1}^{3}+q_{0}q_{1}q_{2}+\tfrac{1}{2}q_{0}q_{1}^{2}+q_{0}^{2}q_{2}+\tfrac{1}{2}q_{0}^{2}q_{1},
\]
with $P_{\text{AC}}=P_{\text{AB}}$ and $Q_{\text{AC}}=Q_{\text{AB}}$.
For the probability that site A is connected to both sites B and C,
we have
\[
P_{\text{ABC}}=p_{2}^{3}+\tfrac{3}{2}p_{1}p_{2}^{2}+\tfrac{3}{4}p_{1}^{2}p_{2}+\tfrac{1}{8}p_{1}^{3},
\]
\[
Q_{\text{ABC}}=q_{2}^{3}+3q_{1}q_{2}^{2}+\tfrac{9}{4}q_{1}^{2}q_{2}+\tfrac{1}{2}q_{1}^{3}+3q_{0}q_{2}^{2}+3q_{0}q_{1}q_{2}+\tfrac{3}{4}q_{0}q_{1}^{2}.
\]
At the percolation threshold, we must have $P_{\text{AB}}=Q_{\text{AB}}$
and $P_{\text{ABC}}=Q_{\text{ABC}}$, and taking into account the
normalization conditions in Eq. (\ref{eq:normal}) we obtain
\begin{equation}
p_{2}+\tfrac{1}{2}p_{1}=1-2\sin\tfrac{\pi}{18}=p_{c}^{(\text{honeycomb)}}\label{eq:pchoneycomb}
\end{equation}
and 
\begin{equation}
q_{2}+\tfrac{1}{2}q_{1}=2\sin\tfrac{\pi}{18}=p_{c}^{(\text{triangular})},\label{eq:pctriangular}
\end{equation}
again in agreement with Eq. (\ref{eq:pc}).

All these predictions show that at least one of the critical percolation points of isotropically directed bonds, when a giant out-going
component (GOUT) is formed, can be simply related to the model of
standard percolation by Eq. (\ref{eq:pc}).  Our numerical results
indicate that the critical point defined by the formation of a GSCC coincides with the formation of
a GOUT. However, as we show in the following section, compared to the GOUT,
the GSCC has a different set of critical exponents.

\section{Critical state of the percolation of isotropically directed bonds}

Redner \citep{Redner1981,Redner1982} and
Dorogovtsev~\citep{Dorogovtsev1982} solved exactly percolation of
directed bonds on a hierarchical lattice obtained by iterating the
process shown in Fig.~\ref{diagrama}. Having the probabilities $p_0$,
$p_1$, and $p_2$ at a given generation of the process, renormalization
group calculations allow one to determine the probabilities
$p^\prime_0$, $p^\prime_1$, and $p^\prime_2$, of the next generation.
The scale invariant states are the fixed points of the renormalization
group. As shown in Fig.~\ref{diagrama}, two of those points represent
the trivial cases of a fully disconnected lattice $p_0=1$ and a fully
connected lattice $p_2=1$. Another fixed point is $p_1=0$ with
$p_2=1/2$ representing the critical scale invariant state for standard
percolation in this hierarchical lattice, while the remaining one,
$p_1=0.49142$ with $p_2=0.25429$, is the critical scale invariant
state for percolation of isotropically directed bonds in this
hierarchical lattice. Surprisingly, this hierarchical lattice has
similarities with the square lattice \citep{Redner1981,Redner1982}, as
it can predict exactly, not only the critical point of standard
percolation, but also the whole critical line $p_2+p_1/2=1/2$. With
the exception of the critical state of standard percolation, all the
other points along the critical line converge through the
renormalization process towards the scale invariant state of
percolation of isotropically directed bonds.


Although these renormalization group calculations do not
yield precise predictions for the correlation-length exponent $\nu$ in $2D$, they give the same value $\nu=\ln_2(13/8)\approx1.428$ for both
standard percolation as well as percolation of isotropically directed
bonds \citep{Redner1981,Redner1982}. Moreover, there are two
order-parameter exponents $\beta_1=0.1342$ and $\beta_2=0.1550$ that are
related to clusters which percolate in a single direction or in both
directions, respectively \citep{Redner1981,Redner1982}. Again here the
method does not obtain exactly the value of the exponent $\beta$ for
$2D$, presented in the next Section, but shows that $\beta_1$ is the same value as the one obtained for
standard percolation in this hierarchical lattice, while $\beta_2$
is shown to be a different exponent. These two different exponents indicate,
at least for this hierarchical lattice, that percolation on isotropically directed bonds is tricritical.

In the next Section, we show that simulation results for
the square, honeycomb and triangular lattices confirm
Eqs. (\ref{eq:pcsquare})--(\ref{eq:pctriangular}). Furthermore, we
show that indeed the fractal dimensions of the two forms of
critical giant clusters, GSCC and GOUT, are different from each other,
but seemingly universal among the different lattices.

\section{Simulation results}

We start by describing our results for two dimensions, while the $3D$
case will be discussed subsequently.  We simulated bidimensional lattices
with linear size ranging from $L=32$ to $L=8192$, taking averages over
a number of samples ranging from $38400$ (for $L=32$) to $150$ (for
$L=8192$), halving the number of samples each time that the linear
size was doubled.  Periodic boundary conditions were employed.
 
Besides checking the predictions for the percolation threshold, our
goal is to obtain the values of the critical exponents associated with
({\it{i}}) clusters which can be traversed in one direction and ({\it{ii}})
clusters which can be traversed in both directions. In the language of
complex networks, these clusters correspond in case ({\it{i}}) to giant
out-components (GOUT) and in case ({\it{ii}}) to the giant strongly-connected
component (GSCC).  For each sample, we identified all the SCCs by
using Tarjan's algorithm~\citep{Tarjan1972}, and  calculated their size
distribution.
At the percolation threshold, we also looked at the giant outgoing
component (GOUT), which corresponds to the GSCC augmented by sites
outside of it which can be reached from those in the GSCC. By
symmetry, the statistical properties of the GOUT must be the same as
those of the giant in-component (GIN), defined as the set of sites not
in the GSCC from which we can reach the GSCC, augmented by sites in
the GSCC itself. Figure \ref{lattice} shows an example of a
square lattice with $L=16$, indicating the GSCC, the GOUT, and the
GIN. Some of the results presented next were obtained with the help
of the Graph Tool software library \citep{peixoto_graph-tool_2014}.

We define the order parameter here as the fraction of sites belonging to
the largest SCC.
For an infinite planar lattice, this order parameter
should behave as 
\begin{equation}
\lim_{L\rightarrow\infty}\frac{\left\langle S\right\rangle }{L^{2}}\sim\left(p-p_{c}\right)^{\beta_{scc}},\label{eq:minf}
\end{equation}
where $\left\langle S\right\rangle$ is the average size of the largest SCC, $p$ is a parameter that controls the distance to the critical
point $p_{c}$, and $\beta_{scc}$ is expected to be a universal critical
exponent. A finite-size scaling ansatz for the order parameter is
\begin{equation}
\frac{\left\langle S\right\rangle }{L^{2}}\sim L^{-\beta_{scc}/\nu}f_{1}\left(\left(p-p_{c}\right)L^{1/\nu}\right),\label{eq:mfss}
\end{equation}
in which $\nu$ is the correlation-length critical exponent and $f_{1}\left(x\right)$
is a scaling function. From this ansatz, we see that precisely at
the critical point we should have
\begin{equation}
\frac{\left\langle S\right\rangle }{L^{2}}\sim L^{-\beta_{scc}/\nu}.\label{eq:mfss0}
\end{equation}
Similarly, we can look at the second moment of the SCC size distribution
(excluding the GSCC), which, for an infinite lattice, should behave
as
\begin{equation}
\lim_{L\rightarrow\infty}\left\langle S^{2}\right\rangle \sim\left(p-p_{c}\right)^{-\gamma_{scc}},\label{eq:chiinf}
\end{equation}
with $\gamma_{scc}$ being another universal critical exponent. The corresponding finite-size
scaling ansatz is 
\begin{equation}
\left\langle S^{2}\right\rangle \sim L^{\gamma_{scc}/\nu}f_{2}\left(\left(p-p_{c}\right)L^{1/\nu}\right),\label{eq:chifss}
\end{equation}
where $f_{2}\left(x\right)$ is also a scaling function, and precisely
at the critical point we should have
\begin{equation}
\left\langle S^{2}\right\rangle \sim L^{\gamma_{scc}/\nu}.\label{eq:chifss0}
\end{equation}

In order to obtain values for these critical exponents, we have to
introduce a parameterization of the probabilities $p_{0}$, $p_{1}$ and
$p_{2}$. We performed two different sets of simulations,
with different parameterizations.  In the first set of numerical
experiments, bonds were occupied with probabilities parameterized as
\begin{equation}
p_{0}=\left(1-p\right)^{2},\quad p_{1}=2p\left(1-p\right),\quad p_{2}=p^{2},\label{eq:ex1}
\end{equation}
with $0\leq p\leq1$, so that, according to Eq. (\ref{eq:pc}), we have
$p=p_{c}$ at the percolation threshold. This corresponds to randomly
assigning on directed bond with probability $p$ on each possible
direction of each pair of nearest neighbors; two opposite directed
bonds between the same pair correspond to an undirected bond.

\begin{figure*}
\subfigure[]{\includegraphics[width=0.33\textwidth]{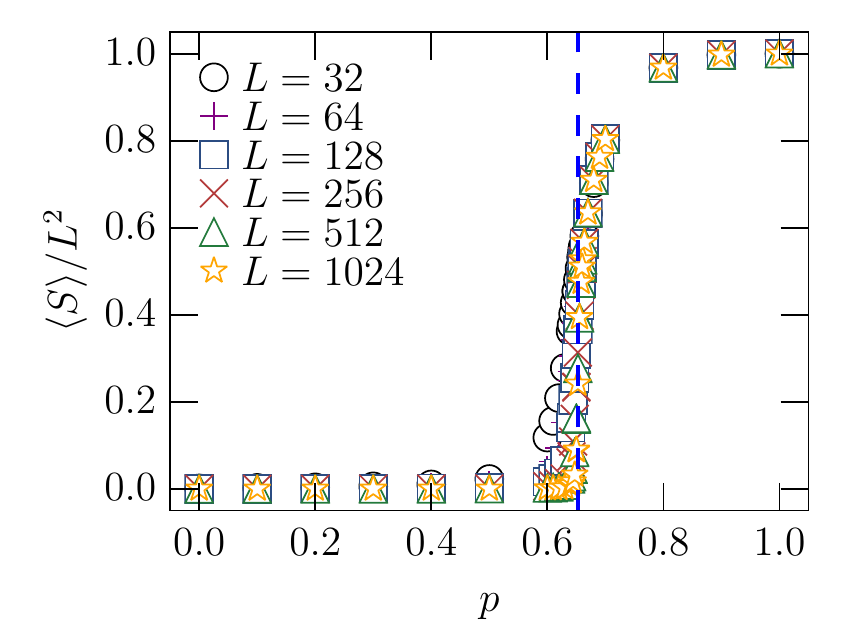}

}\subfigure[]{

\includegraphics[width=0.33\textwidth]{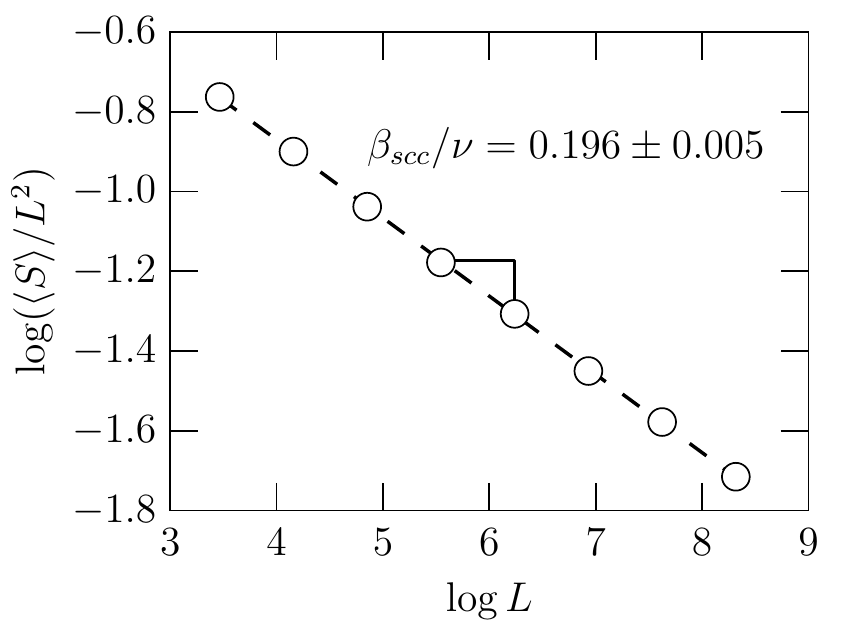}

}\subfigure[]{

\includegraphics[width=0.33\textwidth]{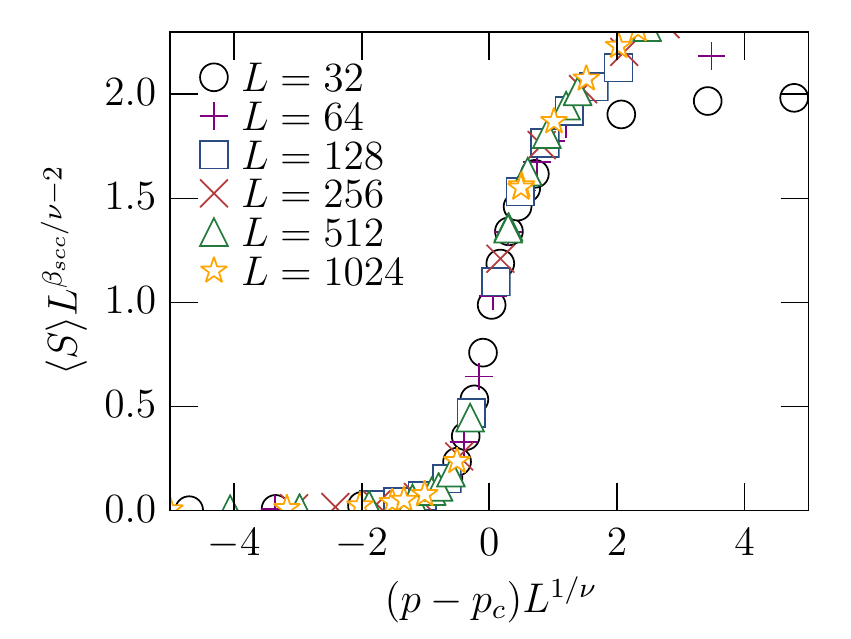}

}\caption{
\label{fig:mexp1}
These results correspond to a honeycomb lattice where each possible
directed bond is occupied with probability $p$, with opposite bonds
between the same pair of sites being present with probability $p^2$,
as parameterized in Eq.~(\ref{eq:ex1}). a) The order parameter for
percolation of isotropically directed bonds defined as the fraction of
nodes in the largest strongly connected component (GSCC). b) At the
critical point the fraction occupied by the GSCC decays as a power
law, yielding the exponent $\beta_{scc}/\nu$.
c) Using the value obtained for
$\beta_{scc}/\nu$, and assuming that the exponent $\nu$ for percolation of isotropically directed bonds is the same as in standard percolation,
$\nu=4/3$, we can collapse all the curves near the critical point.
Here, and in all other plots, error bars are smaller
than the symbols.
}
\end{figure*}

Figure \ref{fig:mexp1} shows results for the SCC order parameter
for honeycomb lattices with sizes ranging from $L=32$ to $1024$.
As depicted in Fig. \ref{fig:mexp1}(a), the threshold probability is
consistent with the result $p\simeq0.653$ predicted by Eq. (\ref{eq:pchoneycomb}).
Figure \ref{fig:mexp1}(b) plots the SCC order parameter at the critical
point, which is expected to scale as in Eq. (\ref{eq:mfss0}), a scaling
form from which we extract $\beta_{scc}/\nu=0.196\pm0.005$. Finally, Fig.
\ref{fig:mexp1}(c) shows a rescaling of the finite-size results according to Eq. (\ref{eq:mfss}). 
%
It is a well known fact~\citep{staufer1994} that percolation has a
single length scale given by the correlation length $\xi$ that
diverges at the critical point as $\xi \sim N^{-\nu}$. Since there is no reason to expect that percolation on isotropically directed
lattices introduces other length scales it is reasonable to assume
that the exponent $\nu$ controlling the scale divergence of SCCs near the critical point is the same as in traditional percolation.  This conjecture is supported by the renormalization group predictions of Redner~\citep{Redner1981} and of Janssen and
Stenull~\cite{Janssen2000,Zhou2012} for the square lattice. Therefore, the best data collapse is obtained assuming for the correlation-length critical exponent the same value as in standard percolation, $\nu = \frac{4}{3}$, which leads to 
\[
\beta_{scc}=0.264\pm0.008.
\]

\begin{figure*}
  \subfigure[]{\includegraphics[width=0.33\textwidth]{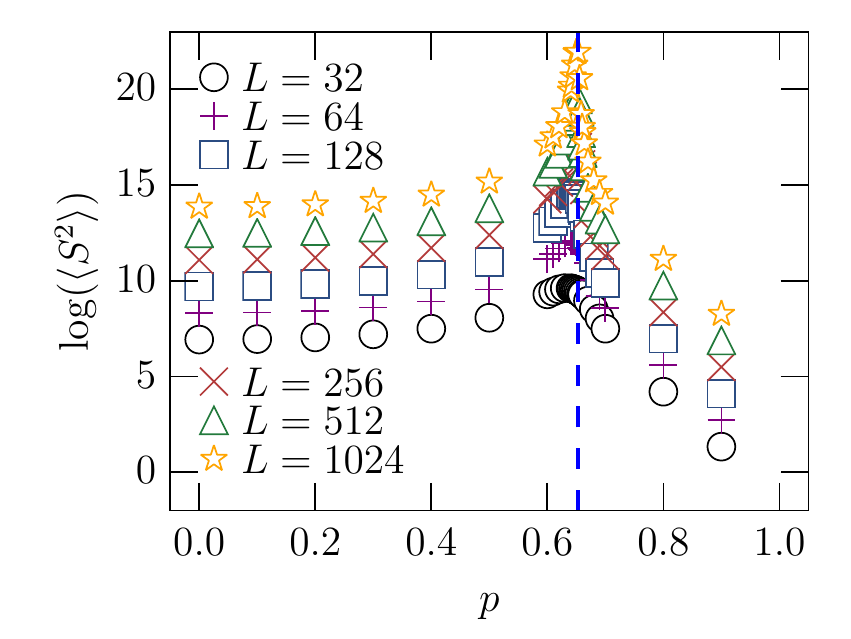}}\subfigure[]{\includegraphics[width=0.33\textwidth]{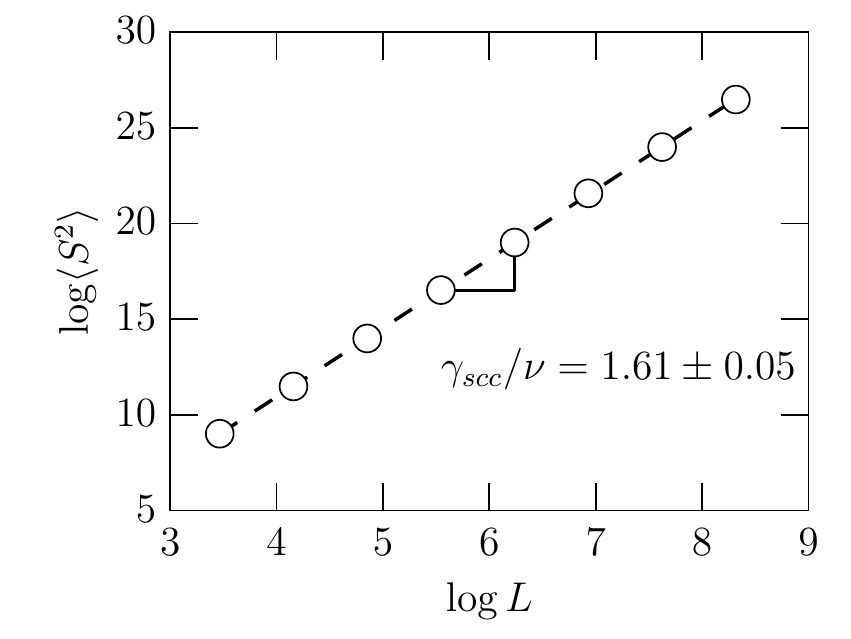}}\subfigure[]{\includegraphics[width=0.33\textwidth]{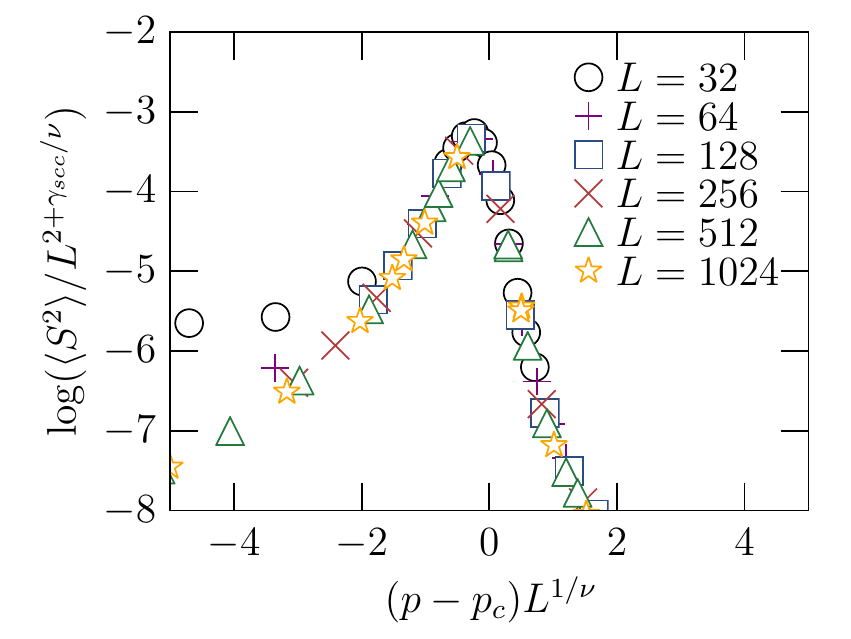}}\caption{\label{fig:chiexp1}
    These results correspond to a honeycomb lattice where each
    possible directed bond is occupied with probability $p$, with
    opposite bonds between the same pair of sites being present with
    probability $p^2$, as parameterized in Eq.~(\ref{eq:ex1}). a) The
    second moment of the distribution of sizes of SCCs, excluding the
    largest SCC.  b) At the critical point the second moment grows as
    a power law, yielding the exponent $\gamma_{scc}/\nu$. c) Using the
    value obtained for $\gamma_{scc}/\nu$, and assuming $\nu=4/3$, we can
    collapse all the curves near the critical point.}
\end{figure*}

For the second moment of the SCC size distribution, Fig. \ref{fig:chiexp1}
shows results for honeycomb lattices. As shown in Fig. \ref{fig:chiexp1}(a),
the value of the percolation threshold is compatible with the prediction
of Eq. (\ref{eq:pchoneycomb}), while from Fig. \ref{fig:chiexp1}(b)
and Eq. (\ref{eq:chifss0}) we obtain $\gamma_{scc}/\nu=1.61\pm0.05$. Again,
the best data collapse of Eq. (\ref{eq:chifss}), shown in Fig. \ref{fig:chiexp1}(c),
is obtained by using $\nu=\frac{4}{3}$, yielding 
\[
\gamma_{scc}=2.15\pm0.07.
\]

\begin{table}
\begin{centering}
\begin{tabular}{cccc}
\hline 
Lattice & $\beta_{scc}$ & $\mathrm{\gamma}_{scc}$ & $d_{scc}$\tabularnewline
\hline 
Triangular & $\,0.264(8)\,$ & $\,2.15(7)\,$ & $\,1.804(5)\,$\tabularnewline
Square & $\,0.261(5)\,$ & $\,2.13(3)\,$ & $\,1.801(8)\,$\tabularnewline
Hexagonal & $\,0.27(1)\,$ & $\,2.15(7)\,$ & $\,1.80(1)\,$\tabularnewline
\hline
\end{tabular}
\par\end{centering}

\caption{\label{T1} Values of the critical exponents
  related to the SCCs, as obtained for the triangular, square and
  hexagonal lattices, for the cases where each possible directed bond
  is occupied with probability $p$, with opposite bonds between the
  same pair of sites being present with probability $p^2$, as
  parameterized in Eq.~(\ref{eq:ex1}). Numbers in parentheses indicate the
  estimated error in the last digit.
}
\end{table}
Table \ref{T1} summarizes the critical exponents obtained under the
first parameterization for the triangular, square, and honeycomb
lattices. We mention that the values obtained for $\beta_{scc}$ and
$\gamma_{scc}$ are all compatible with values extracted from the
simulation results by fitting the data for the largest linear size
with the scaling predictions in Eqs. (\ref{eq:minf}) and
(\ref{eq:chiinf}). We also measured the mass of the GSCC, denoted by
$M_{scc}$, which is predicted to follow
\[
M_{scc}\sim L^{d_{scc}},
\]
with a fractal dimension 
\[
d_{scc}=2-\beta_{scc}/\nu. 
\]
This is confirmed by
the measurements of $d_{scc}$ reported in the last column of Table
\ref{T1}.

In the second set of simulations, bonds were occupied with probabilities
\begin{eqnarray}
p_{0}=&\max\left(0,1-  2p\right),\quad p_{1}=2\min\left(p,1-p\right),\nonumber \\
& p_{2}=\max\left(0,2p-1\right),
\label{eq:ex2}
\end{eqnarray}
again with $0\leq p\leq1$. These probabilities mean that for
$p\leq\tfrac{1}{2}$ there are no undirected bonds, while for
$p\geq\tfrac{1}{2}$ there are no vacancies. Exactly at $p=\frac{1}{2}$
there is a randomly directed bond between each pair of nearest
neighbors. Again, according to Eq. (\ref{eq:pc}), we have $p=p_{c}$ at
the percolation threshold.
The results for the GSCC properties measured under this second parameterization
are compatible with those obtained under the first parameterization.
As shown in Table \ref{tab:secondparam}, the critical exponents $\beta_{scc}$
and $\gamma_{scc}$ and the fractal dimension $d_{scc}$ are all compatible
with the values obtained under the first parameterization.

\begin{table}
\begin{centering}
\begin{tabular}{ccccc}
\hline 
Lattice & $\beta_{scc}$ & $\mathrm{\gamma}_{scc}$ & $d_{scc}$ & $\tau_{scc}$\tabularnewline
\hline 
Triangular & $\,0.26(1)\,$ & $\,2.16(8)\,$ & $\,1.805(8)\,$ & $\,2.07(9)\,$\tabularnewline
Square & $\,0.26(1)\,$ & $\,2.17(4)\,$ & $\,1.802(8)\,$ & $\,2.11(8)\,$\tabularnewline
Hexagonal & $\,0.27(1)\,$ & $\,2.16(7)\,$ & $\,1.80(1)\,$ & $\,2.12(8)\,$\tabularnewline
\hline 
\end{tabular}
\par\end{centering}

\caption{\label{tab:secondparam}Values of the critical exponents
  related to the SCCs, as obtained for the triangular, square and
  hexagonal lattices for the case where undirected bonds appear only
  when all possible vacancies have already been occupied by a directed
  bond, as parameterized by Eq.(\ref{eq:ex2}).  Numbers in parentheses again
  indicate the estimated error in the last digit. Within the error
  bars these values are compatible with those of Table~\ref{T1}.}
\end{table}

\begin{figure}
\includegraphics[width=0.4\textwidth]{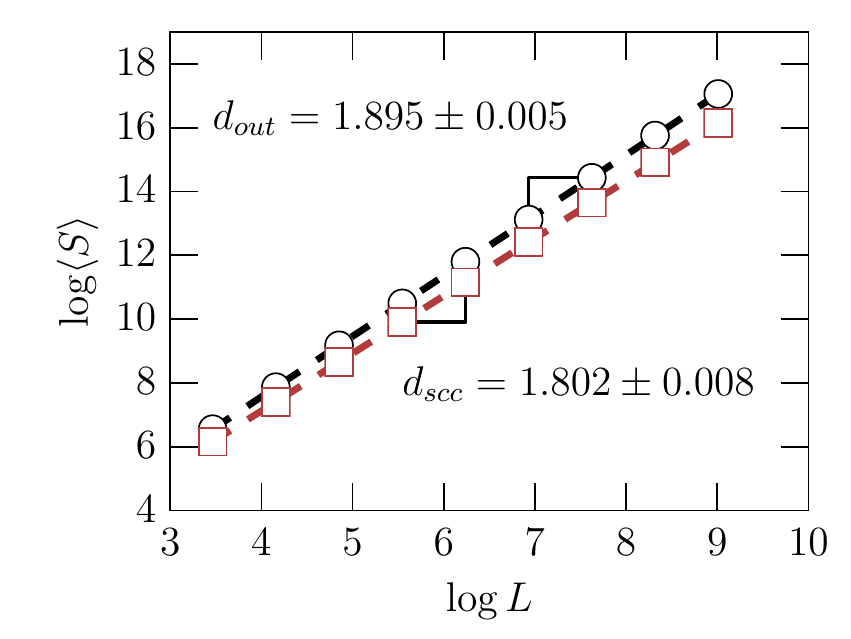}
\caption{\label{fig:dfcompare} 
	The fractal dimensions for the GSCC (squares) and the GOUT
	(circles). These results concern square lattices where all nearest
	neighbors are connected by a directed bond. This corresponds to the
	critical condition when using the parametrization given by
	Eq.(\ref{eq:ex2}). While the fractal dimension of GOUT is compatible
	with that of standard percolation clusters, the GSCCs have a smaller
	fractal dimension.
	}
\end{figure}

Under the second parameterization, besides measuring the mass of the
GSCC associated with the fractal dimension $d_{scc}$, we also measured
the mass of the GOUT, denoted by $M_{out}$, which scales as
\[
M_{out}\sim L^{d_{out}},
\]
where $d_{out}$ is a fractal dimension.
We expect $d_{scc}\leq d_{out}$, as the GSCC is a subset of the
GOUT. Indeed, as shown in Fig.~\ref{fig:dfcompare}, for the
square lattice at the critical point, the fractal dimension $d_{out}$
of the GOUT is compatible with the exact fractal dimension $d_f=91/48$~\citep{staufer1994}
of the critical percolating cluster in standard percolation, while the
value for $d_{scc}$ is about 10\% smaller.

\begin{figure}
\includegraphics[width=0.4\textwidth]{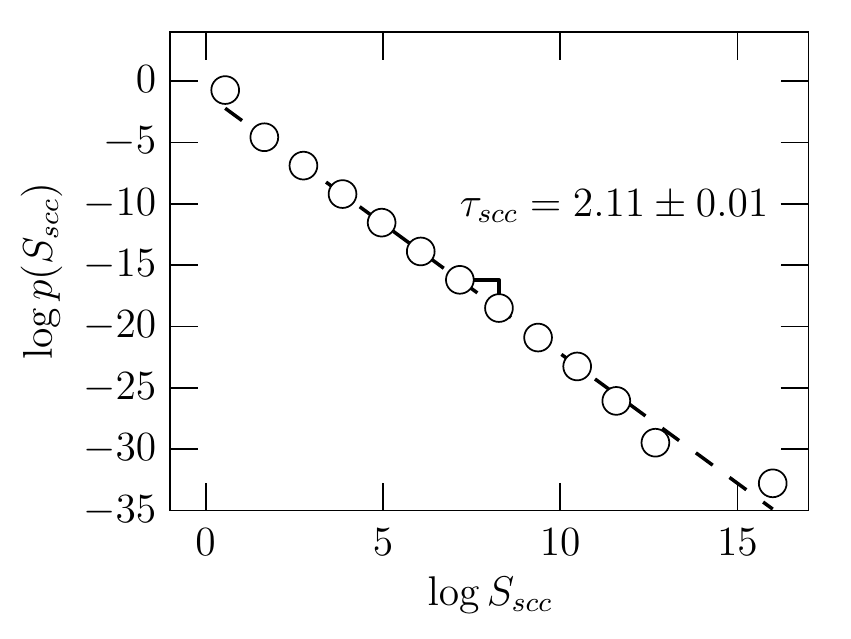}

\caption{\label{fig:gsccprob}Scaling behavior of the SCC size distribution
$p\left(S\right)$ for a square lattice with $L=8196$, under the
second parameterization. }
\end{figure}
\begin{figure}
\includegraphics[width=0.4\textwidth]{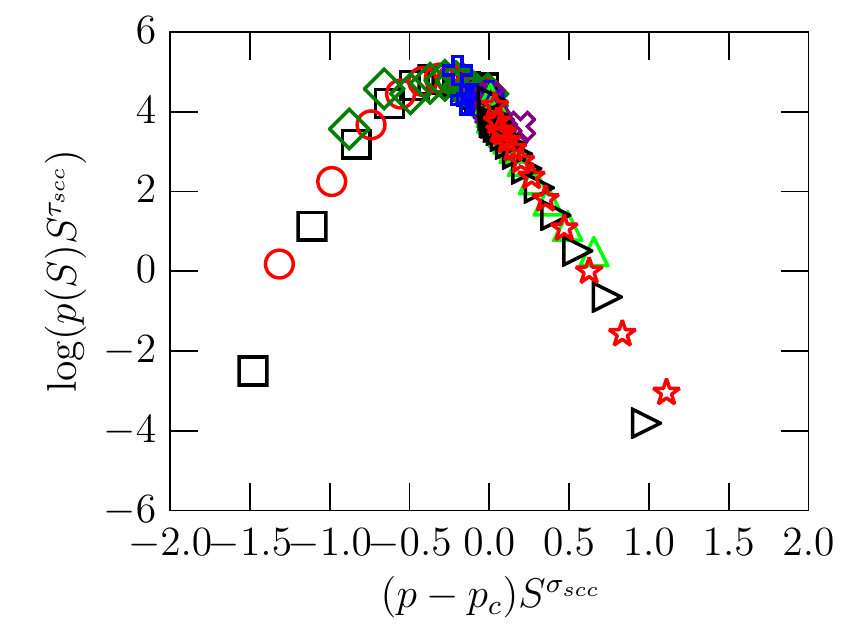}

\caption{\label{fig:gsccps}
		Scaling behavior of the SCC size distribution
		$p\left(S\right)$ for a square lattice with $L=4096$, under the
		second parametrization. The results are for $p=p_c+\delta$ with
		$\delta=-0.01$ (black squares), $\delta=-0.005$ (red circles),
		$\delta=-0.0025$ (green diamonds), $\delta=-0.00075$ (blue cross), $\delta=0.00075$ (violet x), $\delta=0.0025$ (lime up triangles), $\delta=0.005$ (black right triangles) and $\delta=0.01$ (red stars).
		 }
\end{figure}

Finally, we looked at the exponents $\tau_{scc}$ and $\sigma_{scc}$ associated
with the SCC size distribution, expected to scale as
\begin{equation}
p\left(S\right)\sim S^{-\tau_{scc}}f_{3}\left(\left(p-p_{c}\right)S^{\sigma_{scc}}\right),\label{eq:ps}
\end{equation}
where $f_{3}\left(x\right)$ is yet another scaling function. The
Fisher exponent $\tau_{scc}$ associated with the scaling behavior of the
SCC size distribution $p\left(S\right)$ at the critical point is
defined as
\[
p\left(S\right)\sim S^{-\tau_{scc}}.
\]
Figure \ref{fig:gsccprob} shows $p\left(S\right)$ as a function
of $S$ for a square lattice with linear size $L=8196$ at the percolation
threshold. The value $\tau_{scc}=2.11\pm0.08$ obtained is compatible with
the scaling relation $\tau_{scc}=2+\beta_{scc}/\left(\beta_{scc}+\gamma_{scc}\right)$. Values
of $\tau_{scc}$ for the three lattices are reported in the last column
of Table \ref{tab:secondparam}. On the other hand, Fig. \ref{fig:gsccps}
shows rescaled plots of $p\left(S\right)$ for a triangular lattice
with $L=1000$, exhibiting good data collapse based on Eq. (\ref{eq:ps})
with $\tau_{scc}=2.11$ and $\sigma_{scc}=0.414$, in agreement with the scaling
relation $\sigma_{scc}=1/\left(\beta_{scc}+\gamma_{scc}\right)$.

\begin{figure*}
	\subfigure[]{\includegraphics[width=0.33\textwidth]{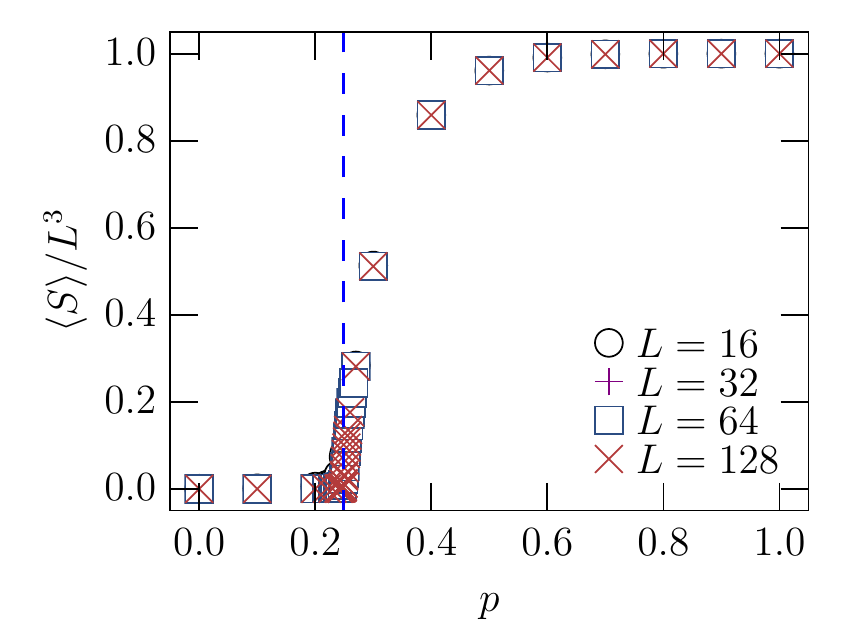}
		
	}\subfigure[]{
	
	\includegraphics[width=0.33\textwidth]{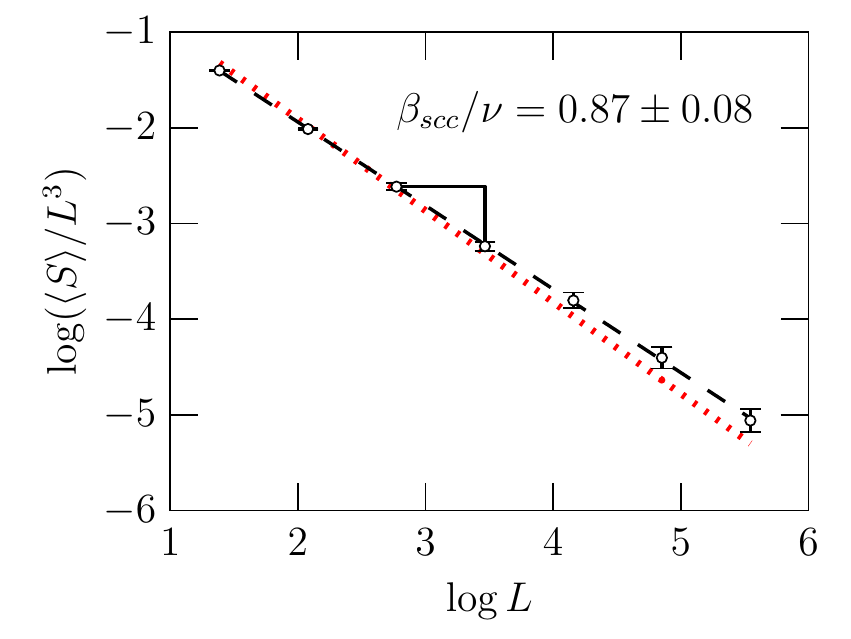}
	
}\subfigure[]{

\includegraphics[width=0.33\textwidth]{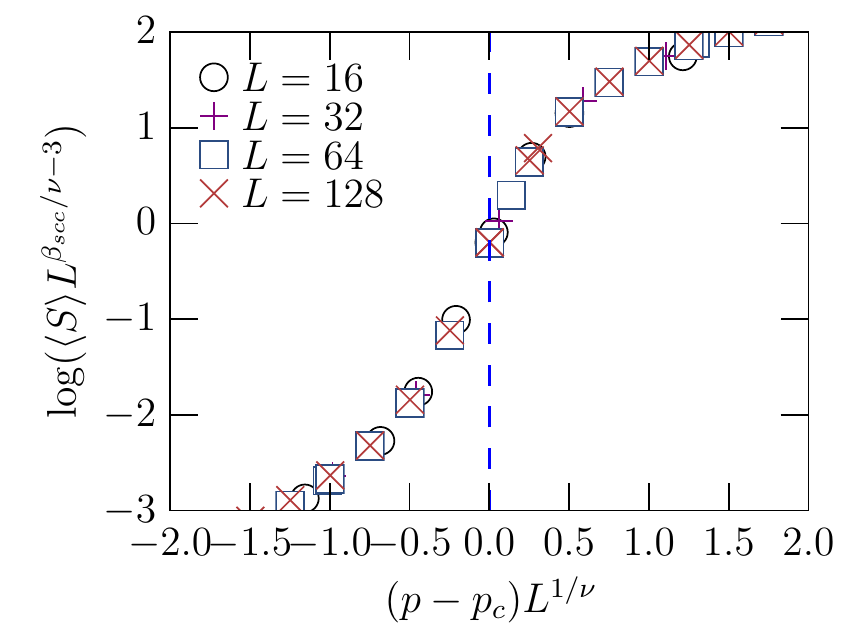}

}\caption{
\label{fig:3d1}
These results correspond to a cubic lattice where each possible
directed bond is occupied with probability $p$, with opposite bonds
between the same pair of sites being present with probability $p^2$,
as parameterized in Eq.~(\ref{eq:ex1}). a) The order parameter for percolation of isotropically directed bonds, corresponding to the fraction occupied by the largest SCC.  b) At the critical point, the
fraction occupied by the largest cluster decays as a power law,
yielding the exponent $\beta_{scc}/\nu$. The red dotted line
corresponds to the form $L^{-2\beta/\nu}$, with $\beta=0.418$ and
$\nu=0.876$ being the exponents for standard percolation in 3D. Given the error bar of the points and the possibility
of finite size deviations, our results allow for the possibility that in three dimensions
$\beta_{scc}=2\beta$.  c) Using the value obtained for
$\beta_{scc}/\nu=0.87$, and assuming that the exponent $\nu=0.876$ as in
standard percolation, we can collapse all the curves near the
critical point.}
\end{figure*}

\begin{figure*}
	\subfigure[]{\includegraphics[width=0.33\textwidth]{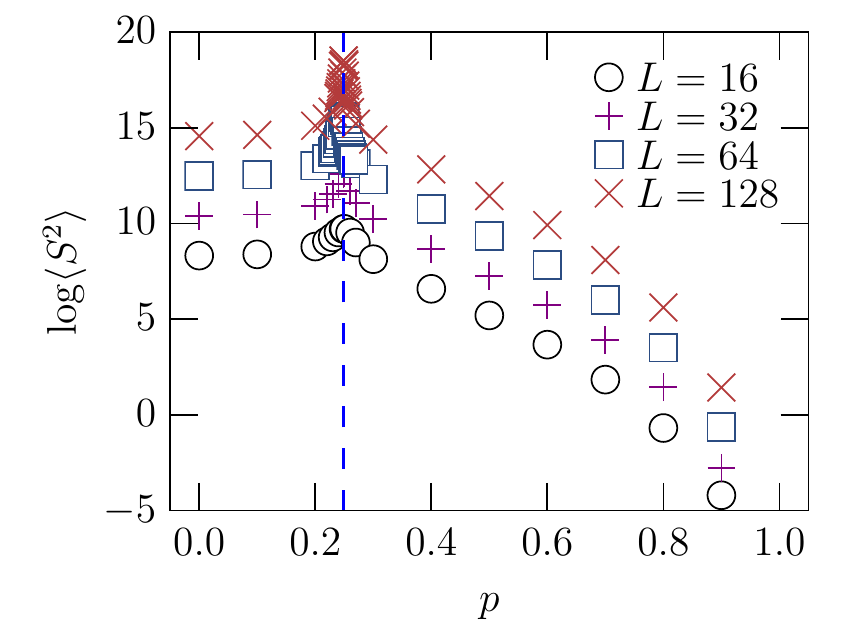}
		
	}\subfigure[]{
	
	\includegraphics[width=0.33\textwidth]{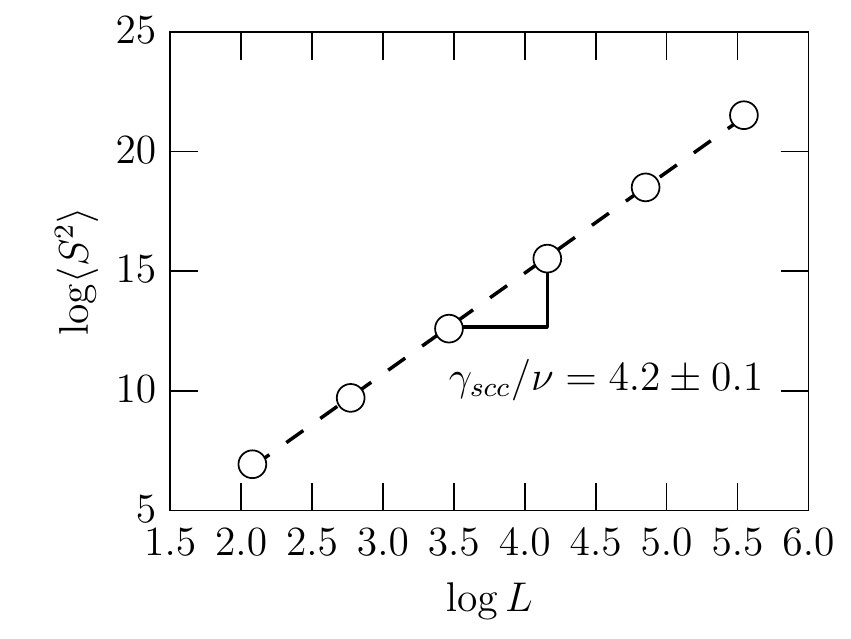}
	
}\subfigure[]{

\includegraphics[width=0.33\textwidth]{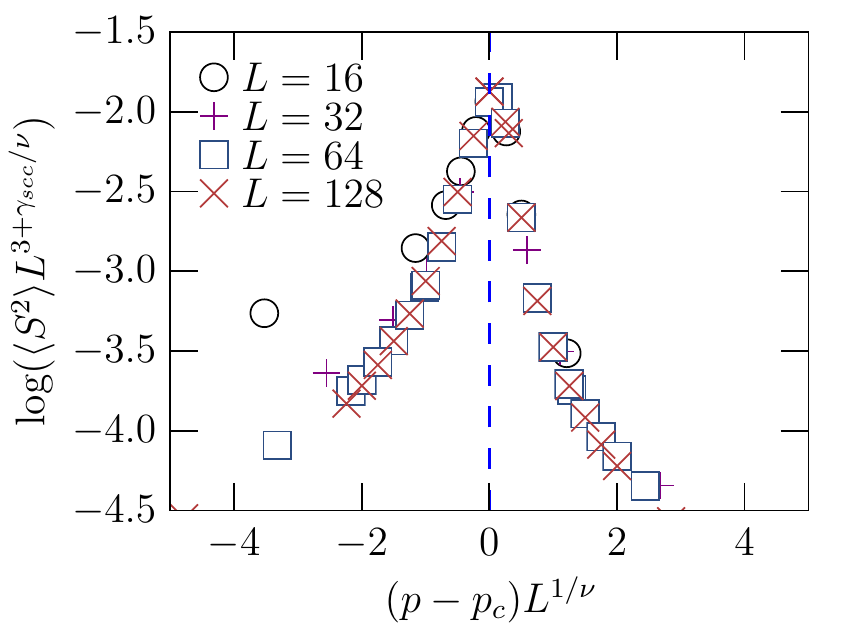}

}\caption{
          \label{fig:3d2}
These results correspond to a cubic lattice where each possible directed
bond is occupied with probability $p$, with opposite bonds between the
same pair of sites being present with probability $p^2$, as
parameterized in Eq.~(\ref{eq:ex1}). a) The second moment of the
distribution of sizes of SCCs, excluding the largest SCC.  b) At the
critical point, the second moment grows as a power law, yielding the
exponent $\gamma_{scc}/\nu$. c) Using the value obtained for $\gamma_{scc}/\nu$,
and assuming $\nu=0.876$ as in standard percolation, we can collapse all the curves near the critical point.}
\end{figure*}

\begin{figure}
	\includegraphics[width=0.4\textwidth]{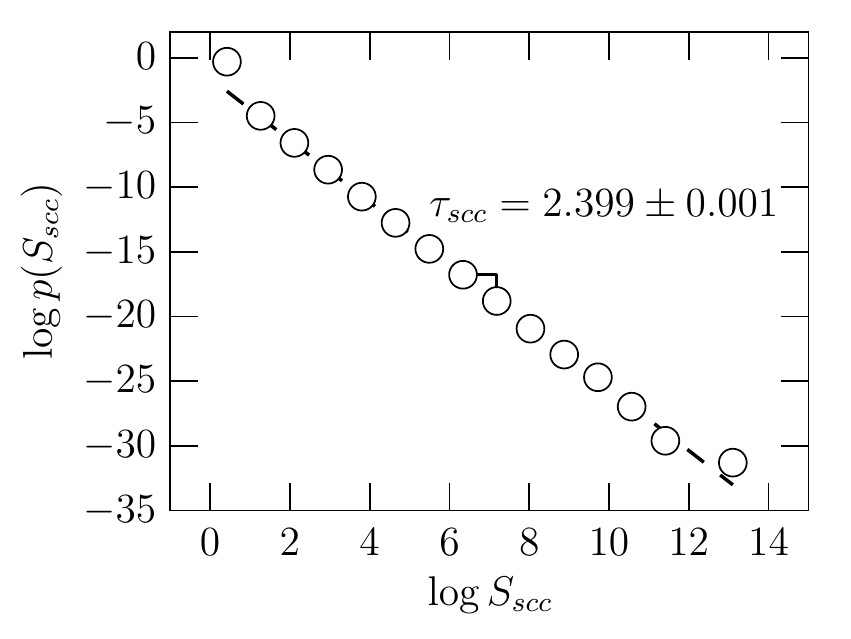}
	
	\caption{
	\label{fig:tau3}Scaling behavior of the SCC size distribution
	$p\left(S\right)$ for a cubic lattice with $L=128$,
	under the second parameterization. }
\end{figure}

Finally, we have also performed simulations on a cubic lattice under the first
parameterization, Eq.~(\ref{eq:ex1}). We simulated lattices
 with linear size going from $L=16$ to $128$, taking averages over a number of samples going from $9600$ ($L=16$) to $1000$ ($L=128$).
Fig.~\ref{fig:3d1} shows results concerning
the order parameter, while Fig.~\ref{fig:3d2} shows results concerning the second
moment of the distribution of sizes of SCCs. As in the case of two
dimensions, the critical point has the same value as for standard
percolation $p_c=0.2488$ ~\cite{Wang2013,Lorenz1998}. At the critical point, both
quantities scale as power laws, yielding the exponents
$\beta_{scc}/\nu$ and $\gamma_{scc}/\nu$. Assuming that the exponent
$\nu$ is the same as in standard percolation, $\nu=0.876$ ~\cite{Ballesteros1999,Hu2014},
 we have $\beta_{scc}=0.76(1)$ and $\gamma_{scc}=3.6(1)$. As we show in
figs.~\ref{fig:3d1} and ~\ref{fig:3d2}, the curves for different system sizes can be collapsed
using these values for the exponents. Figure~\ref{fig:tau3} shows the
distribution of sizes of SCCs for cubic lattices with $L=128$. The value
obtained for the Fisher exponent $\tau_{scc}=2.40\pm0.01$ is, within error bars,
consistent with the hyperscaling relation $\tau_{scc}=1+3/d_{scc}$, with $d_{scc}=3-\beta_{scc}/\nu\approx2.13\pm 0.01$.

\section{Discussion}

We investigated the percolation of isotropically directed bonds,
and presented a conjectured
expression for the location of the percolation threshold, which we
showed to be exact for the square, triangular and honeycomb lattices. 

We have also performed extensive computer simulations and investigated the
percolation properties of the strongly-connected components (SCC), the
out-components (OUT) and the in-components (IN). Contrary to what happens
in directed scale-free networks \citep{Schwartz2002}, on the regular
lattices considered in this paper the percolation threshold is the
same for SCCs, OUTs and INs. This is related to the fact that, once we
are slightly above $p_{c}$, there is an infinite number of paths (in
the thermodynamic limit) connecting the opposite sides of the
lattice. We also obtain an apparently universal order-parameter
exponent for the SCCs that is larger (or, equivalently, a fractal
dimension which is smaller) than the one for both the OUTs and the
INs. Moreover, the exponents obtained for the giant out-components are
the same as those obtained for standard
percolation~\citep{Zhou2012}. This is in agreement with an approximate
real-space renormalization group prediction \cite{Redner1981} that the order-parameter
exponents are different for clusters which can be traversed only in one direction and for clusters which can be traversed in both
directions. Also, simulations in a cubic lattice
allowed us to confirm that the critical point for this case also coincides with that of standard percolation. Finite-size scaling
for this case shows that the exponent $\nu$ is the same
as that of standard percolation, while the exponents $\beta_{scc}$ and $\gamma_{scc}$
for the giant strongly connected component in percolation of isotropically directed bonds differ from those
of standard percolation.

Note that the value of the order-parameter exponent obtained for the
SCCs from Figs~\ref{fig:mexp1}(b) and~\ref{fig:chiexp1}(b) is also distinct from the value of the GOUT order-parameter
exponent obtained in Refs. \citep{Inui1999,Janssen2000} as a function
of the anisotropy introduced by allowing a preferred direction. In
that case, the exponent is simply given by the product of a crossover
exponent and the usual GOUT exponent of standard percolation.

The correlation function gives the probability that two sites
separated by a distance $r$ belong to the same cluster and, at the
critical transition, decays for large distances r as $g(r)\sim
r^{-2\beta/\nu}$ ~\cite{Fisher1974,KimNicholas2005}.  In the case of
percolation of directed bonds, different correlation functions can be
defined. Here we define $g_{out}(r)$ as the probability that a given
node is in the out-component of another node separated by a distance
$r$. Alternatively, we define $g_{scc}(r)$ as the probability that two
sites separated by a distance $r$ belong to the same SCC. Assuming
that finding a path in one direction or the other are uncorrelated
events, we have $g_{scc}(r)=g_{out}(r)\times g_{in}(r)$. Since the
in/out components are in the same universality class as standard
percolation, we have that, considering uncorrelated events, the value
of $\beta_{scc}$ should be twice that for standard percolation.  If in
fact these events are correlated, one could expect $\beta_{scc}/2$ to
be smaller than the exponent $\beta$ of standard percolation. In
standard percolation, a cutting bond~\citep{Coniglio1982}
is a bond that if removed results in the loss of connection in a
cluster.  In our extension, a directed bond can be a cutting bond in
each direction or possible in both directions, we call this later case
a double cutting bond. The presence of double cutting bonds should lead
to correlations between the connectivity events in the opposite
directions. Note that the same event (including/removing this cutting bond)
would determine the presence or not of a path from one side to the
other in both directions. In standard percolation the density of
cutting bonds decays as $L^{1/\nu-d}$~\citep{Coniglio1982}.
Considering that being a cutting bond in each direction are
independent events, the density of these double cutting bonds should
be the square of the density of cutting bonds in standard percolation
$L^{2/\nu-2d}$.  Since this density decreases faster than $L^{-d}$, the
number of such double cutting bonds should be zero in large enough
lattice sizes, indicating that no correlation should be observed.  In
the case of two dimensions this relation is true within the error
bars, $\beta_{scc}=0.27\pm 0.01\approx 2\times 5/36=0.2777$.  In the
case of three dimensions, the obtained value for $\beta_{scc}=0.76\pm
0.08$ is smaller than expected, as the value of standard percolation
is $\beta=0.418\pm0.001$~\cite{Ballesteros1999}. However, this
deviation is still within the error bars and may also arise from finite-size
effects.


\begin{acknowledgments}
The thank the Brazilian agencies CNPq, CAPES, FUNCAP, NAP-FCx, the National Institute of Science and Technology for Complex Fluids (INCT-FCx) and the National Institute of Science and Technology for Complex Systems (INCT-SC) in Brazil for financial support.
\end{acknowledgments}

\clearpage
\bibliographystyle{apsrev4-1}
\bibliography{random_diode}

\end{document}